\RequirePackage[l2tabu, orthodox]{nag}
\RequirePackage{snapshot}

\documentclass[9pt,twocolumn]{extarticle}
\makeatletter
\if@twocolumn
  \usepackage[dvips,letterpaper,top=0.5in, bottom=0.5in, left=0.75in, right=0.5in,includefoot,heightrounded]{geometry}
\else
  \usepackage[dvips,letterpaper,margin=1in,includefoot,heightrounded]{geometry}
\fi

\usepackage{srcltx}

\usepackage{amsmath}
\usepackage{amssymb,amsfonts}

\usepackage{abstract}

\usepackage{graphicx}
\usepackage[usenames,dvipsnames]{color}
\usepackage[usenames,dvipsnames]{xcolor}

\usepackage{booktabs}

\usepackage{setspace}
\usepackage{flushend}

\usepackage{url}
\usepackage{algorithm}
\usepackage{algorithmic}

\usepackage{threeparttable}

\usepackage[short,12hr]{datetime}

\usepackage{hyperref}
\usepackage[hyperpageref]{backref}

\usepackage{fontawesome}
\usepackage{amsrefs}

\newtheorem{definition}{Definition}

\newtheorem{conjecture}{Conjecture}
\newtheorem{example}{Example}

\makeatletter

\newcommand{\printtitle}{%
\makeatletter
\if@twocolumn

\twocolumn[%
  \maketitle
  \begin{onecolabstract}
    \myabstract
  \end{onecolabstract}
  \begin{center}
    \small
    \textbf{Keywords}
    \\\medskip
    \mykeywords
  \end{center}
  \bigskip
]
\saythanks
\else
  \maketitle
  \begin{onecolabstract}
    \myabstract
  \end{onecolabstract}
  \begin{center}
    \small
    \textbf{Keywords}
    \\\medskip
    \mykeywords
  \end{center}
  \bigskip
  \onehalfspacing
\fi
\makeatother
}

\title{A New Algorithm for Double Scalar Multiplication over Koblitz Curves}

\author{%
Jithra Adikari$^\ast$%
\and
Vassil S. Dimitrov%
\thanks{%
Department of Electrical and Computer Engineering,
University of Calgary, Calgary, AB, Canada.
Email:~\protect\url{jithra.adikari@gmail.com},\url{vdvsd103@gmail.com}}
\and
Renato~J.~Cintra%
\thanks{%
Signal Processing Group,
Departamento de Estat\'istica,
Universidade Federal de Pernambuco,
Recife, PE, Brazil.
E-mail:~\protect\url{rjdsc@de.ufpe.br}}
}

\date{}

\newcommand{\myabstract}{%
Koblitz curves are a special set of elliptic curves and have improved performance in computing scalar multiplication in elliptic curve cryptography due to the Frobenius endomorphism. Double-base number system approach for Frobenius expansion has improved the performance in single scalar multiplication. In this paper, we present a new algorithm to generate a sparse and joint $\tau$-adic representation for a pair of scalars and its application in double scalar multiplication. The new algorithm is inspired from double-base number system. We achieve 12\% improvement in speed against state-of-the-art $\tau$-adic joint sparse form.
}

\newcommand{\mykeywords}{%
Elliptic curve cryptography, Koblitz curves, field programmable gate array, $\tau$-adic joint sparse form.
}

\begin{document}

\printtitle

\section{Introduction}

Let Koblitz curve be
\begin{equation}
E_a : y^2 + xy = x^3 + ax + 1,
\end{equation}
where $a \in \{ 0, 1\}$, $E(\mathbb{F}_{2^m})$ a group of points on $E_a$ for some extension field $\mathbb{F}_{2^m}$ and $n$ the group order of $E_a(\mathbb{F}_{2^m})$. Any point $P \equiv (x, y) \in E_a(\mathbb{F}_{2^m})$ has following properties
\begin{equation}
\label{eq:fm}
[\tau]P = (x^2, y^2)
\qquad
\text{and}
\qquad
-P = (x, x + y),
\end{equation}
where $\tau$ is called the Frobenius map over $E_a(\mathbb{F}_{2^m})$. Further, there exists a point at infinity denoted by $\mathcal{O}$~\cite{koblitz_cm_91}. The point at infinity satisfies the properties,
\begin{equation}
\label{eq:fm0}
[\tau]\mathcal{O} = \mathcal{O}
\qquad
\text{and}
\qquad
-\mathcal{O} = \mathcal{O}.
\end{equation}

The Frobenius mapping of a point can be computed by squaring its coordinates. The cost of the squaring is very cheap and fast in its hardware implementation with both polynomial and normal basis representations~\cite{hanmenvan04}. In the digital signature verification of an elliptic curve cryptosystem, double scalar multiplication $[k]P + [l]Q$ consumes most computational power, where $P, Q \in E_a(\mathbb{F}_{2^m})$ and $k, l \in [1, n - 1]$. The scalar $k, l$ are represented in $\tau$-adic expansion to obtain the advantage of Frobenius map by replacing point doublings.

Let $\mathbb{Z}[\tau]$ be a ring of polynomials in the form $\sum_{i=0}^{l-1}u_i\tau^i$, where $l$ is the length of the polynomial, $u_i \in \{0,\pm 1\}$ for all $0 < i < l - 1$ and $u_{l-1} = \pm1$. First, both scalars are converted or reduced in $\mathbb{Z}[\tau]$ to complex numbers such that $l$ is minimal. The reduction in $\mathbb{Z}[\tau]$ is defined as $\rho \equiv k\mod{\delta}$, where $k$ is an integer in $[1, n - 1]$ and $\delta = (\tau^m - 1)/(\tau - 1)$. Next $\tau$-adic non-adjacent form $\sum_{i=0}^{l-1}u_i\tau^i$ with $l/3$ average Hamming weight is computed~\cite{solinas00}. Then two $\tau$NAFs are used as inputs to generate $\tau$-adic joint sparse form ($\tau$JSF) of both scalars with average Hamming weight of $l/2$~\cite{ciet_improved_2003}.

Dimitrov~\emph{et~al.} has introduced the two dimensional Frobenius expansion (TDFE) $\sum_{i=0}^{l - 1}\sum_{j=0}^{k}u_{i,j}\tau^i(\tau - 1)^j$, where $l$ is the length of the $\tau$-adic expansion, $u_{i,j} \in \{0, \pm1\}$ and $k$ is an integer to compute single scalar multiplication~\cite{dimitrov2008provably}. Note that TDFE can be reduced to a polynomial in $\mathbb{Z}[\tau]$~\cite{dimitrov2008provably}.

Our approach towards the double scalar representation is based on TDFE. Our algorithm delivers a joint and sparse two dimensional representation that can be reduced to $\mathbb{Z}[\tau]$. It is used with Straus\rq~idea~\cite{strauss64} to compute double scalar multiplication and perform minimum number of point additions. Our new algorithm, joint two dimensional Frobenius expansion (JTDFE) is having 15\% improvement in terms of speed compared to $\tau$JSF in its implementation on a field programmable gate array (FPGA).

This paper is arranged as follows: two dimensional Frobenius expansion is discussed in Section~\ref{sec:preliminaries}. The construction of new algorithm JTDFE is discussed in Section~\ref{sec:blk:algo}. Section~\ref{sec:implandresult} explains the hardware implementation. We conclude the paper in Section~\ref{sec:conclusion}.

\section{Two Dimensional Frobenius Expansion}
\label{sec:preliminaries}
The Frobenius map $\tau$ is a complex number with value $(\mu + \sqrt{-7})/2$, where $\mu = (-1)^{1 - a}$. A complex number in the form of $a + \tau b$, where $a, b \in \mathbb{Z}$ is called a Kleinian integer~\cite{conway_quaternions_2003}. Next, we define the \{$\tau, \tau - 1$\}-Kleinian integer.

\begin{definition}{\{$\tau, \tau - 1$\}-Kleinian integer}
{A Kleinian integer $\omega$ of the form $\omega = \pm\tau^x(\tau - 1)^y$, where $x,y \in \mathbb{Z}^*$ is called a \{$\tau, \tau - 1$\}-Kleinian integer.}
\end{definition}

The two dimensional Frobenius expansion of an integer can be represented as in the following equation:
\begin{equation}
\label{eq:2dfe}
k = \sum_{i=1}^{d}s_i\tau^{a_i}(\tau - 1)^{b_i},
\end{equation}
where $d$ is the length of the expansion $s_i = \pm1$ and $a_i, b_i \in \mathbb{Z}^*$. We rearrange~\eqref{eq:2dfe} as follows:
\begin{equation}
\label{eq:2dfe_rearr}
k = \sum_{l = 1}^{\mathrm{max}(b_i)}(\tau - 1)^l\left (\sum_{i=1}^{\mathrm{max}(a_{i,l})}s_{i,l}\tau^{a_{i,l}}\right ),
\end{equation}
where $\mathrm{max}(a_{i,l})$ is the maximum power of $\tau$ that is multiplied by $(\tau - 1)^l$ in~\eqref{eq:2dfe_rearr}.

Algorithm~\ref{alg:2dfe} illustrates the routine to compute the single scalar multiplication $[k]P$ when the \{$\tau, \tau - 1$\}-expansion of $k$ is given. In order to simplify, we denote the terms corresponding to $(\tau - 1)^l$ in~\eqref{eq:2dfe_rearr} with $r_l(k)$, i.e. $r_l(k) = \sum_{i=1}^{\mathrm{max}(a_{i,l})}s_{i,l}\tau^{a_{i,l}}$. The multiplication $[\tau - 1]P$ costs one Frobenius mapping and a point addition. Therefore, $\mathrm{max}(b_i)$ should be limited when the two dimensional Frobenius expansion is computed.

\begin{algorithm}
  \caption{Scalar Multiplication using Two Dimensional Frobenius Expansion}
  \label{alg:2dfe}
  \begin{algorithmic}[1]
    \REQUIRE Two dimensional Frobenius expansion of $k \in \mathbb{N}$ and a point $P \in E(\mathbb{F}_{2^m})$.
    \ENSURE $Q = [k]P$.
    \STATE $P_0 \gets P$
    \STATE $Q \gets \mathcal{O}$
    \FOR {$l = 0$ to $\mathrm{max}(b_i)$}
      \STATE $S \gets [r_l(k)]P_l$\quad \{one dimensional $\tau$NAF corresponding to $(\tau - 1)^l$ in~\eqref{eq:2dfe_rearr}\}
      \STATE $P_{l+1} \gets [\tau]P_l - P_l$
      \STATE $Q \gets Q + S$
    \ENDFOR
    \STATE \textbf{return} ($Q$).
  \end{algorithmic}
\end{algorithm}

Finding an algorithm that returns a fairly short representation of $k$ as the sum of \{$\tau, \tau - 1$\}-Kleinian integers is an absolute need. The greedy algorithm given in~\cite{dimitrov2008provably} is used to obtain such a representation. Greedy algorithm does not always return the canonical \{$\tau, \tau - 1$\}-expansion. Note that the complexity of the greedy algorithm depends crucially on the time spent to find the closest \{$\tau, \tau - 1$\}-Kleinian integer to the current Kleinian integer.

However, finding the closest Kleinian integer  in an intermediate step of  greedy algorithm is achieved by precomputing all Kleinian integers $\pm\tau^x(\tau - 1)^y$ for $x, y \in \mathbb{Z}^*$ less than certain bounds and using an exhaustive search. Using divide-and-conquer principle, Dimitrov~\emph{et~al.} have invented an effective method to generate an efficient two dimensional Frobenius expansion for computing single scalar multiplication~\cite{dimitrov2008provably}. Further they have conjectured following:

\begin{conjecture}{Length of Two Dimensional Frobenius Expansion}
\label{cnj:len_ki}
Every Kleinian integer $\xi = a + b\tau$, can be represented as the sum of at most
$O\left(\log{N(\xi)}/\log{\log{N(\xi)}}\right)$
\{$\tau, \tau - 1$\}-Kleinian integers, where $N(\xi) = (a + b\tau)(a + b\overline{\tau})$ is the norm of $\xi$.
\end{conjecture}

They highlighted that use of two complex bases has increased the theoretical difficulties in proving the Conjecture~\ref{cnj:len_ki}. Nevertheless, that lead to a more important practical blocking algorithm given in~\cite{dimitrov2008provably}.

\section{Joint Blocking Algorithm}
\label{sec:blk:algo}
In this section, we present the construction of our new algorithm to return a joint and sparse representations for a pair of Kleinian integers $\eta_0, \eta_1 \in \mathbb{Z}[\tau]$. Algorithm~\ref{alg:new_jblk} illustrates the procedure to compute a joint two dimensional Frobenius expansion in $\mathbb{Z}[\tau]$ for a pair of Kleinian integers.

\begin{algorithm}
  \caption{Blocking Algorithm Computing Joint Two Dimensional Frobenius Expansion}
  \label{alg:new_jblk}
  \begin{algorithmic}[1]
    \REQUIRE A pair of Kleinian integers $\eta_0, \eta_1 \in \mathbb{Z}[\tau]$, window size $w$ and precomputed table of optimal joint two dimensional Frobenius expansions for all possible pairs of Kleinian integers $\sum_{i=0}^{w-1}u_{0,i}\tau^i$ and $\sum_{i=0}^{w-1}u_{1,i}\tau^i$.
    \ENSURE A pair of lists $L_0, L_1$ of \{$\tau, \tau - 1$\}-Kleinian integers representing $(\eta_0, \eta_1)$.
    \FOR {$i = 0$ to $1$}
    	\STATE $L_i \gets \emptyset$
	\STATE compute $\tau$-adic expansion $\eta_i = \sum_{j=1}^{l}\eta_{i,j}\tau^j$, where $\eta_{i,j} \in \{0,1\}$\label{step:three}
    \ENDFOR
    \FOR {$i = 0$ to $\lfloor l/w\rfloor$}
	\STATE find optimal joint two dimensional Frobenius expansion of pair of $\sum_{j=0}^{w-1}u_{0,j+iw}\tau^j$ and $\sum_{j=0}^{w-1}u_{1,j+iw}\tau^j$\label{step:oj2dfe}
	\STATE multiply each term by $\tau^{iw}$ and add to $L_0$ or $L_1$\label{step:seven}
	\STATE $i \gets i + 1$
    \ENDFOR
    \STATE \textbf{return} $(L_0, L_1)$.
  \end{algorithmic}
\end{algorithm}

A window size $w$ is fixed prior to running the algorithm. Then the optimal joint two dimensional Frobenius expansions for all possible pairs of $w$-bit $\tau$-adic representations are precomputed and given as another input.

First, two $\tau$-adic expansions $\sum_{i=0}^{l-1}u_i\tau^i$, where $u_i \in \{0, 1\}$ and $l$ is the length of the longer expansion, in $\mathbb{Z}[\tau]$ are computed. Next both $\tau$-adic expansions are arranged as in~\eqref{eq:jte} to generate \textit{joint columns}.
\begin{equation}
\label{eq:jte}
\begin{pmatrix}
\eta_0\\
\eta_1
\end{pmatrix} = \begin{pmatrix}
\eta_{0,l-1} & \ldots & \eta_{0,1} & \eta_{0,0}\\
\eta_{1,l-1} & \ldots & \eta_{1,1} & \eta_{1,0}
\end{pmatrix}
\end{equation}
The $i^\text{th}$ joint column in~\eqref{eq:jte} has two elements $\eta_{0,i}, \eta_{1,i} \in \{0, 1\}$ for all $i$ satisfying $0 \leq i < l$. If one $\tau$-adic expansion is shorter than the other, then the coefficients of higher degrees of $\tau$ of shorter expansion should be set to zero.

Two $\tau$-adic expansions are separated into $w$-bit $\left\lceil l/w\right\rceil$ number of blocks. The least significant $w$ bits of $\tau$-adic expansion have the label block 0, while the most significant bits have label block $\left\lfloor l/w\right\rfloor$.

At step~\ref{step:oj2dfe} of Algorithm~\ref{alg:new_jblk}, $i^{\text{th}}$ block of $\eta_0$ and $\eta_1$ representations are considered to find the $i^{\text{th}}$ block of optimal joint two dimensional Frobenius expansion. This is achieved by a look-up-table approach. Once the $i^{\text{th}}$ block of optimal joint two dimensional Frobenius expansion is obtained, all elements are multiplied by $\tau^{iw}$ and appended to the relevant lists. We repeat this step for $\left\lceil l/w\right\rceil$-times to obtain the complete joint two dimensional Frobenius expansion. Example~3 illustrates the execution of Algorithm~\ref{alg:new_jblk}.

\begin{figure*}
\centering
\includegraphics[width=1.5\columnwidth]{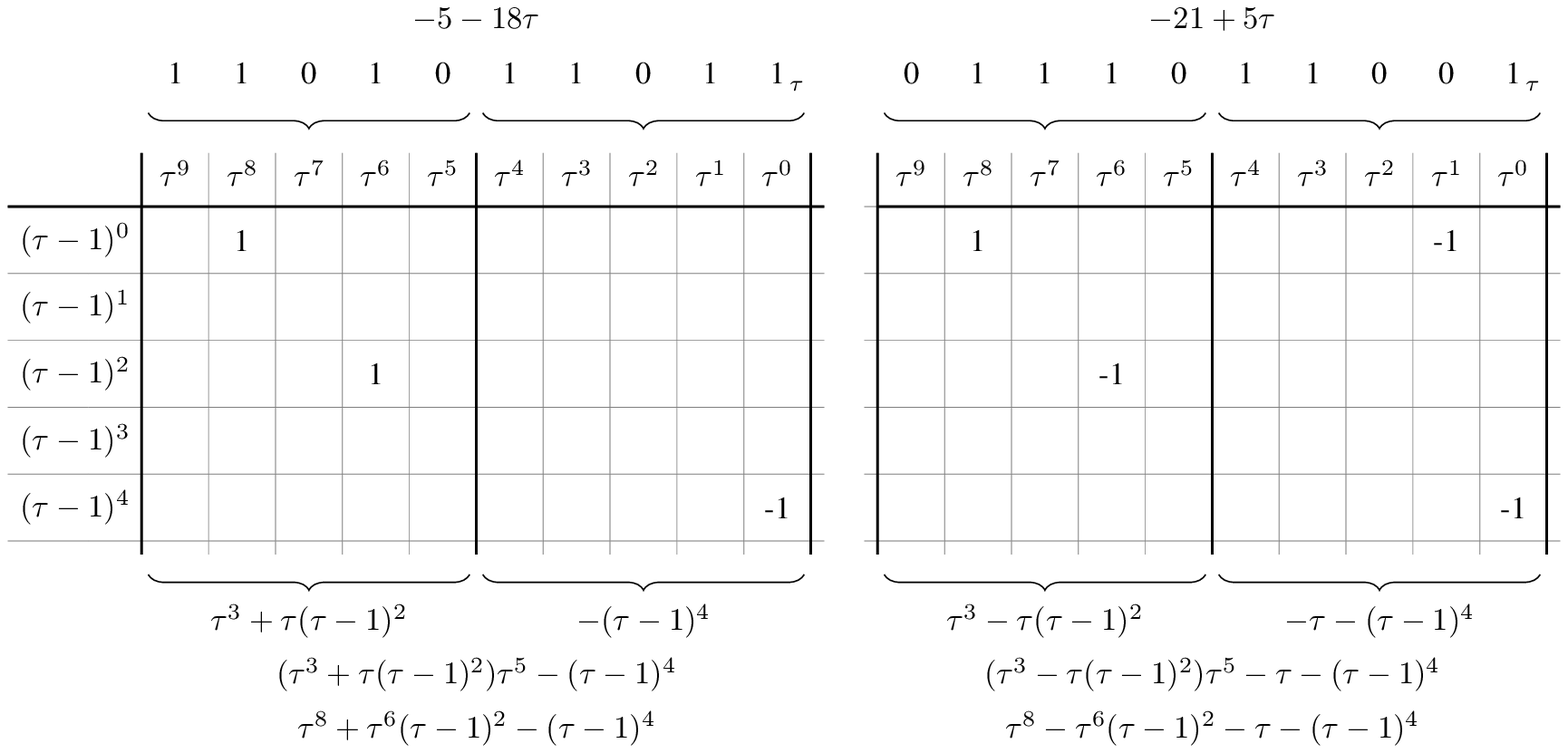}
\caption{Graphical Representation of Joint Two Dimensional Frobenius Expansion of $\eta_0 = -5 - 18\tau$ and $\eta_1 = -21 + 5\tau$}
\label{fig:tau_rep}
\end{figure*}

\begin{example}{Joint Two Dimensional Frobenius Expansion}
{\label{ex:jtdfe}
We consider two Kleinian integers $\eta_0 = -5 - 18\tau$ and $\eta_1 = -21 + 5\tau$ with $a = 1$ in this example. As the first step we compute $\tau$-adic expansions for both $\eta_0$ and $\eta_1$ (Step~\ref{step:three} of Algorithm~\ref{alg:new_jblk}):
\[
\begin{array}[c]{rcrccccccccccl}
-5 - 18\tau & = & ( & 1 & 1 & 0 & 1 & 0 & 1 & 1 & 0 & 1 & 1 & )_{\tau}\\
-21 + 5\tau & = & ( &  & 1 & 1 & 1 & 0 & 1 & 1 & 0 & 0 & 1 & )_{\tau}
\end{array}
\]
To construct the joint expansion we need to make both expansions in the same length. Therefore we append a zero to the beginning of the $\tau$-adic representation of $-21 + 5\tau$:
\[
\begin{array}[c]{rcrccccccccccl}
-21 + 5\tau & = & ( & 0 & 1 & 1 & 1 & 0 & 1 & 1 & 0 & 0 & 1 & )_{\tau}
\end{array}
\]
Let $w = 5$. We divide both $\tau$-adic expansions into $5$-bit blocks and find the optimal joint two dimensional Frobenius expansion for each block. The pair $11011_\tau$ and $11001_\tau$ have $-(\tau - 1)^4$ and $-\tau - (\tau - 1)^4$. Most significant bits pair have $\tau^3 + \tau(\tau - 1)^2$ and $\tau^3 - \tau(\tau - 1)^2$ (Step~\ref{step:oj2dfe} of Algorithm~\ref{alg:new_jblk}). We multiply last pair by $\tau^5$ to obtain final results (Step~\ref{step:seven} of Algorithm~\ref{alg:new_jblk}). The optimal joint two dimensional Frobenius expansion is given by:
\[
\begin{pmatrix}
-5 - 18\tau\\
-21 + 5\tau
\end{pmatrix} = \left(
\begin{array}[c]{c}
\tau^8~+~\tau^6(\tau - 1)^2\hspace{26pt}-~(\tau - 1)^4 \\
\tau^8~-~\tau^6(\tau - 1)^2~-~\tau~-~(\tau - 1)^4
\end{array}
\right)_{\text{J2DFE}}
\]}
\end{example}
The main advantage of a joint representation in double scalar multiplication in elliptic curve cryptography is that Straus' method can be applied with some precomputations to improve the efficiency~\cite{strauss64}. Considering Example~3, if $A = P + Q$ and $S = P - Q$ are precomputed, we can compute $[- 5 - 18\tau]P + [- 21 + 5\tau]Q$ with four point additions. We do not consider any additions due to $(\tau - 1)$ terms in the total cost.
\begin{multline}
\label{eq:cal_point}
[- 5 - 18\tau]P + [- 21 + 5\tau]Q = [\tau^8]A + [\tau^6(\tau - 1)^2]S\\ -[\tau]Q - [(\tau - 1)^4]A
\end{multline}
Then we can apply Algorithm~\ref{alg:2dfe} to compute final point with~\eqref{eq:cal_point}. The point negations over Koblitz curves is only a field addition and can be neglected in terms of cost compared to field multiplication. Fig.~\ref{fig:tau_rep} illustrates the graphical representation of joint two dimensional Frobenius expansion of $\eta_0 = -5 - 18\tau$ and $\eta_1 = -21 + 5\tau$.

The generation of precomputed optimal joint representations for all possible combinations of pairs of Kleinian integers for a given window $w$ is achieved by an exhaustive search. This computation needs to be done only once per curve and a given window size. %

\section{Hardware Implementation and Results}
\label{sec:implandresult}
\begin{figure*}
\centering
\includegraphics[width=1.5\columnwidth]{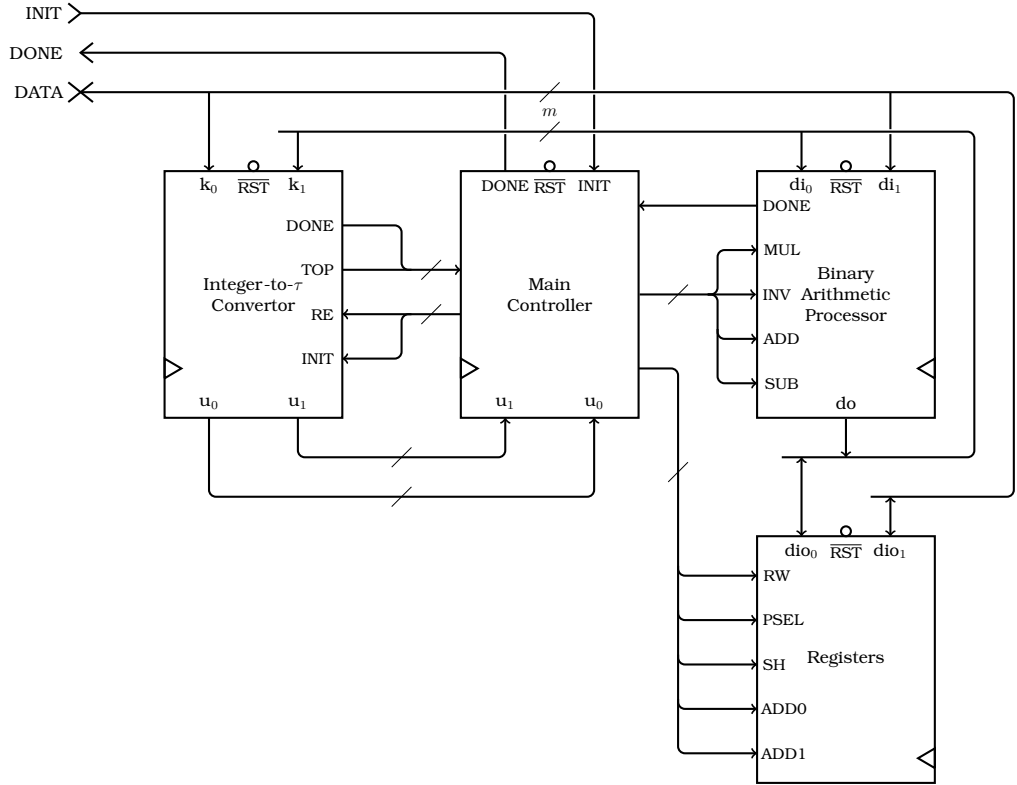}
\caption{Top-Level Components of Circuit for Double Scalar Multiplication with Joint Two Dimensional Frobenius Expansion}
\label{fig:tau_tdfe}
\end{figure*}

The double scalar multiplication over $\mathbb{F}_{2^{163}}$ with joint two dimensional Frobenius expansion is implemented in VHDL and placed and routed to Xilinx \texttt{XC4VLX200} FPGA by executing Xilinx Integrated Software Environment (ISE\texttrademark ) version \texttt{9.2i}. The window size is set to $w=5$ and maximum exponent of $\tau - 1$ is limited to four. We describe the hardware architecture of our circuit in this section. The top-level design components and architecture of the circuit are illustrated in Fig.~\ref{fig:tau_tdfe}.

The circuit is partitioned into four high-level components, namely, main controller (MC), binary arithmetic processor (BAP), integer-to-$\tau$ converter (ITC) and registers. In our implementation, databus width is set to 163 bits. Other than $u_0$ and $u_1$, inputs and outputs of main controller are handshaking signals between MC and other units.

The binary arithmetic processor performs four basic arithmetic operations needed for point multiplication, namely, addition, squaring, multiplication and inversion. All arithmetic operations are performed in the normal basis representation. Addition and squaring can be executed in a single clock cycle. Addition is an exclusive OR ($\mathrm{XOR}$) operation and squaring is a cyclic shift operation in the normal basis representation~\cite{agnew_arithmetic_1993}. Multiplication is a direct implementation of Massey-Omura multiplier with computing four bits in one clock cycle~\cite{patent_4587627}. Therefore we need only forty one clock cycles for the multiplication. The inversion is performed with Itoh-Tsuji architecture~\cite{Itoh_fast_1988}. It needs nine multiplications to calculate the inversion of an element in $\mathbb{F}_{2^{163}}$. Once the multiplication or inversion is performed, binary arithmetic processor sends out a job completion signal by setting DONE of BAP to high.

The primary job of the integer-to-$\tau$ converter is to compute the joint two dimensional Frobenius expansion from a pair of integers. Our implementation comprises of two integer-to-$\tau$ converters with lazy reduction introduced in~\cite{brumley10}, because it is faster and needs less area in hardware implementations. The converter is slightly modified to generate nonnegative elements for the $\tau$-adic expansion, whereas the circuit proposed in~\cite{brumley10} generates the $\tau$NAF. Then a precomputed look-up-table is used to compute joint two dimensional Frobenius expansion. The signal DONE of ITC is high when first $w$ bits of each expansion is available for processing.

The registers are used to store point coordinates and intermediate values during point additions. Further some registers can perform cyclic shift operation to facilitate Frobenius mapping on points $P$, $Q$, $P+Q$, and $P-Q$.

The main controller is designed with a finite state machine to perform the double scalar multiplication with other three components. With the INIT of MC set to high, main controller begins loading integers $k_0, k_1$ and $P$, $Q$ point coordinates $x_P$, $y_P$, $x_Q$, $y_Q$ to registers. Then $k_0$ and $k_1$ are loaded into the integer-to-$\tau$ converter simultaneously. Once DONE of ITC is high, the joint two dimensional Frobenius expansion is read to main controller and the double scalar multiplication is started. The main controller knows that it has reached to the end of computation, when the TOP of ITC is high. Final results are stored in the registers and DONE of MC is set to high.

Affine coordinates are used in precomputations. Mixed coordinates are used for computing $P$ and $Q$ related calculations and needs 8~field multiplications and 5~field squarings. For other point additions, i.e. $P \pm Q$ related computations we have used L\'{o}pez-Dahab projective coordinates in this implementation~\cite{lopez_fast_1999}. These point additions require 13~field multiplications and 4 field squarings.

The hardware implementations are carried out for both $\tau$-adic joint sparse form and joint two dimensional Frobenius expansion based double scalar multiplication. A window value $w = 5$ and maximum $\tau - 1$ exponent $\mathrm{max}(b_i) = 4$ are selected for the joint two dimensional Frobenius expansion implementation. The $y^2 + xy = x^3 + x + 1$ is considered over binary field $\mathbb{F}_{2^{163}}$. We have considered the curve parameters and field for implementation which are specified by NIST. We have implemented both circuits in Xilinx \texttt{XC4VLX200} FPGA and tested for 10,000 pair of integers, $k$, $l$ and pair of points, $P$, $Q$. The summary of the experimental results are given in Table~\ref{tab:jtdfe}.

\begin{table*}
\renewcommand{\arraystretch}{1.3}
\caption{Experimental Results for $\tau$JSF and J2DFE based double scalar multiplication over $\mathbb{F}_{2^{163}}$}
\vspace{5pt}
\label{tab:jtdfe}
\centering
\begin{threeparttable}
\begin{tabular}{cccccc}
\toprule
& Max. & Area & Increase & Av. time & Gain in \tabularnewline
Algo. & Freq. & (Num. of & in area & per calc. & time \tabularnewline
& (MHz) & slices) & (\%)\textsuperscript{a} & ($\mu$s) & (\%)\textsuperscript{b} \tabularnewline
 \midrule
$\tau$JSF & 75.364 & 9,217 & - & 479.609 & - \tabularnewline
J2DFE & 76.559 & 13,403 & 45.42 & 418.675 & 12.70\tabularnewline
\bottomrule
\end{tabular}
\begin{tablenotes}
\item [(a)] Increase in area is given against $\tau$JSF.
\item [(b)] Gain in time is given against $\tau$JSF.
\end{tablenotes}
\end{threeparttable}
\end{table*}

The results collected in Table~\ref{tab:jtdfe} are based on the synthesis goals set for speed maximization. Time is read for each algorithm when the circuit is operating at its maximum frequency.

\paragraph{Note:}
\emph{Timings given for single scalar multiplication in~\cite{dimitrov2008provably} are very smaller than the figures for double scalar multiplication presented in this paper. That is mainly due to three reasons: Firstly, the clock speed in~\cite{dimitrov2008provably} is two times as fast as that of this implementation. Secondly, average number of point additions in double scalar multiplication is more than twice of that in single scalar multiplication. Thirdly, field multiplication in~\cite{dimitrov2008provably} needs 9 clock cycles, while in this implementation we need 41 clock cycles.}

\section{Conclusions}
\label{sec:conclusion}
The joint two dimensional Frobenius expansion outperforms the state-of-the-art $\tau$-adic joint sparse form in double scalar multiplication over Koblitz curves, in speed, according to the experimental results presented in Table~\ref{tab:jtdfe}. The area of the new architecture has increased by about 45\% of that of $\tau$JSF architecture. Having greater values for window sizes and maximum $\tau - 1$ exponents, the speed of the double scalar multiplication can be improved. When the window size is increased the size of look-up-table in integer-to-$\tau$ conversion grows exponentially. We will investigate on different combinations of $w$ and maximum $\tau - 1$ exponent as future work.


\begin{thebibliography}{1}


\bibitem{koblitz_cm_91}
N.~Koblitz, ``{CM-Curves with Good Cryptographic Properties},'' in \emph{CRYPTO
  '91: Proceedings of the 11th Annual International Cryptology Conference on
  Advances in Cryptology}.\hskip 1em plus 0.5em minus 0.4em\relax London, UK:
  Springer-Verlag, 1992, pp. 279--287.

\bibitem{hanmenvan04}
D.~Hankerson, A.~Menezes, and S.~Vanstone, \emph{Guide to Elliptic Curve
  Cryptography}.\hskip 1em plus 0.5em minus 0.4em\relax Springer, 2004.

\bibitem{solinas00}
J.~A. Solinas, ``Efficient arithmetic on Koblitz curves,'' \emph{Design Codes
  Cryptography}, vol.~19, no. 2-3, pp. 195--249, 2000.

\bibitem{ciet_improved_2003}
M.~Ciet, T.~Lange, F.~Sica, and J.~Quisquater, ``Improved algorithms for
  efficient arithmetic on elliptic curves using fast endomorphisms,'' in
  \emph{Advances in Cryptology, {EUROCRYPT} 2003}, 2003, pp. 388--400.

\bibitem{dimitrov2008provably}
V.~S. Dimitrov, K.~U. J{\"a}rvinen, M.~J. Jacobson, W.~F. Chan, and Z.~Huang,
  ``Provably sublinear point multiplication on Koblitz curves and its
  hardware implementation,'' \emph{Computers, {IEEE} Transactions on}, vol.~57,
  pp. 1469--1481, 2008.

\bibitem{strauss64}
E.~G. Straus, ``Addition chains of vectors (problem 5125),'' \emph{American
  Mathematical Monthly}, vol.~70, pp. 806--808, 1964.

\bibitem{conway_quaternions_2003}
J.~H. Conway and D.~Smith, \emph{On Quaternions and Octonions}, 1st~ed.\hskip
  1em plus 0.5em minus 0.4em\relax {AK} Peters, 2003.

\bibitem{brumley10}
B.~B. Brumley and K.~U. J{\"a}rvinen, ``Conversion algorithms and
  implementations for Koblitz curve cryptography,'' \emph{Computers, IEEE
  Transactions on}, vol.~59, pp. 81--92, 2009.

\bibitem{agnew_arithmetic_1993}
G.~Agnew, T.~Beth, R.~Mullin, and S.~Vanstone, ``Arithmetic operations in
  ${GF}(2^m)$,'' \emph{Journal of Cryptology}, vol.~6, no.~1, pp. 3--13, Mar.
  1993.

\bibitem{patent_4587627}
J.~K. Omura and J.~L. Massey, ``Computational method and apparatus for finite
  field arithmetic,'' no. 4587627, May 1986, (US Patent 4587627).

\bibitem{Itoh_fast_1988}
T.~Itoh and S.~Tsujii, ``A fast algorithm for computing multiplicative inverses
  in ${GF}(2^m)$ using normal bases,'' \emph{Information and Computation},
  vol.~78, no.~3, pp. 171--177, Sept. 1988.

\bibitem{lopez_fast_1999}
J.~L\'opez and R.~Dahab, ``{Fast Multiplication on elliptic curves Over
  $GF(2^m)$ without precomputation},'' in \emph{Cryptographic Hardware and
  Embedded Systems}, ser. Lecture Notes in Computer Science.\hskip 1em plus
  0.5em minus 0.4em\relax Springer Berlin / Heidelberg, 1999, vol. 1717, pp.
  316--327.

\end{thebibliography}
\end{document}